# REAL-TIME DETECTION OF GRAVITATIONAL MICROLENSING

M.R. PRATT , C. ALCOCK , R.A. ALLSMAN , D. ALVES , T.S. AXELROD ,
A. BECKER, D.P. BENNETT , K.H. COOK , K.C. FREEMAN ,
K. GRIEST , J. GUERN , M. LEHNER , S.L. MARSHALL , B.A. PETERSON ,
P.J. QUINN , D. REISS , A.W. RODGERS , C. STUBBS , W. SUTHERLAND AND
D. WELCH

*Department Astronomy,*
*University of Washington, Seattle, WA 98195*

*Center for Particle Astrophysics, Berkeley, CA 94720*

*Lawrence Livermore National Laboratory, Livermore, CA 94550*

AND

*Mt. Stromlo and Siding Spring Observatories,*
*Australian National University, Weston, ACT 2611, Australia*

( The MACHO Collaboration )

**Abstract.** Real-time detection of microlensing has moved from proof of concept in 1994 [10, 2] to a steady stream of events this year. Global dissemination of these events by the MACHO and OGLE collaborations has made possible intensive photometric and spectroscopic followup from widely dispersed sites confirming the microlensing hypothesis [4]. Improved photometry and increased temporal resolution from followup observations greatly increases the possibility of detecting deviations from the standard point-source, point-lens, inertial motion microlensing model. These deviations are crucial in understanding individual lensing systems by breaking the degeneracy between lens mass, position and velocity. We report here on GMAN (Global Microlensing Alert Network), the coordinated followup of MACHO alerts.

1. **Introduction**

The MACHO project is engaged in an ongoing, time resolved survey of stars in the Magellanic Clouds and galactic bulge to search for microlensing by intervening compact objects [1] [7, 3]. The apparent amplification of a lensed star varies over time with the relative motion of star, lens and observer. This is given by

$$A = \frac{u^2 + 2}{u\sqrt{u^2 + 4}} \quad \text{with} \quad u(t) = \sqrt{u_{min}^2 + \left(\frac{2(t - t_{peak})}{\hat{t}}\right)^2} \quad (1)$$

where $u = b/R_E$ is the impact parameter in units of the Einstein radius

$$R_E = \sqrt{\frac{4GmD_{OL}D_{LS}}{D_{OS}c^2}}. \quad (2)$$

Here $D_{OL}, D_{LS}$ & $D_{OS}$ are distances between observer, source and lens and $m$ is the mass of the lens. The timescale of an event is given by $\hat{t} = 2R_E/v_\perp$, where $v_\perp$ is the relative perpendicular motion of the unperturbed line of sight and lens. This simple functional form is based on point–source and point–mass approximations and an assumption of uniform linear motion between observer, source and lens. In this regime only $\hat{t}$ provides useful information about the lensing system.

It is possible to extract additional information about the lensing system if one can detect deviations from this simple form. These include:

- microlensing parallax due to the accelerated motion of earth throughout the event
- resolution of the finite size of the source star by the lens
- lensing of binary or coincident sources
- lensing by binary or planetary systems.

At least three of these have been detected [11, 5, 1]. Both parallax and finite–source models provide additional information about the lens velocity, breaking some of the degeneracy between the physical parameters making up $\hat{t}$. These effects are present at low levels in a significant fraction of events but are usually undetectable with the present photometric precision and time coverage. For example, caustic crossing events and planetary perturbations can produce significant flux changes on timescales of an hour. The survey systems in current operation are designed to maximize the number of detected events and are not well suited for monitoring single events at this rate. As there are numerous, under-subscribed, small aperture telescopes with instrumentation adequate for microlensing followup it is not desirable to divert the survey instruments to monitor individual events in progress. Rather it is beneficial to coordinate observations of both the survey system and followup telescopes; using all available data to assist in prioritizing observations. In addition, with the development of more sophisticated triggers, the followup system may become an integral part of the survey in the verification of events.

---

[1] see also Stubbs, *et al.* and Bennett, *et al.* , this volume

## 2. Global Microlensing Alert Network

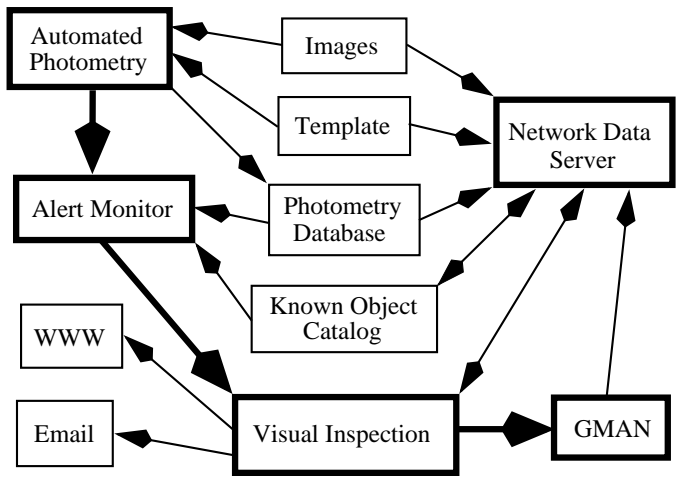

Figure 1. Information flow in the MACHO alert system.

Global Microlensing Alert Network (GMAN) is the coordinated followup of MACHO microlensing alerts. The process by which events are identified and scheduled is shown in Figure 1. The two–color images of each night's survey data are processed in real–time at Mount Stromlo Observatory. This is accomplished with SoDophot, a special purpose photometry code that makes use of a template image taken in good conditions to "warm–start" the photometry. If a star has varied significantly from its "template" value it is reported as anomalous and an automated analysis is performed. Stars which have $7\sigma$ high-points in both passbands, pass a variety of data quality cuts, and are not listed as variables in the "Known Object Catalog" are reported to collaboration members together with best fit microlensing parameters. Another automated process downloads full light curves of the reported objects to each collaborator's local computer for interactive viewing and analysis. This process produces on the order of ten low level alerts per day. Although aided by the automated analysis, the process of identifying microlensing events is ultimately done by a human after close inspection of the lightcurve and template image. The event is then posted to the MACHO Alert WWW page [2] and email is sent to subscribers of the MACHO alert email service. [3] The star is also scheduled for observation at GMAN observatories.

Telescopes participating in GMAN are

- MSO 30inch, Australia: priority to microlensing followup
- CTIO 0.9m, Chile: roughly 1 hour of service observing every night
- UTSO 0.61m, Chile: 50 full nights of service observing in 1995
- Wise Observatory 1.0m, Israel: several hours per week
- Mt. John 24inch, New Zealand: priority to microlensing followup.

GMAN images are processed at the site where the are acquired. Photometry is accomplished with IRAF scripts calling DaophotII [9]. These data are stored on site as photometry files from individual images. Photometry is usually finished within 12 hours of image acquisition and available to team members to aid in subsequent scheduling. Normalization is performed on-the-fly by the "Network Data Server" using a list of reference stars obtained from the MACHO data set.

Figure 2 shows the MACHO data on the 26th alert of the 1995 season together with the GMAN data taken at the CTIO 0.9m telescope. This event was identified approximately 10 days before peak, allowing good coverage throughout the peak from the GMAN site. In most cases, the GMAN measurements exceed the the precision of the survey data and are therefore more likely to expose irregularities in the lightcurve.

A more spectacular demonstration of the success of GMAN is the first real–time detection of an unusual event in progress. This event, shown in Figure 3, was found to deviate from standard microlensing significantly before the time of the first peak. The object was then placed at the highest priority in the GMAN observing schedule and a spectrum was obtained at the AAT. It is likely that this event is either a binary source, coincident source or binary lens. Fits to multiple source or multiple lens systems obviously require many additional parameters and the multiple lens fit in particular is notorious for it's convergence problems. The additional GMAN data should greatly improve the quality and reliability of these fits. The data in Figure 3 have been normalized to unit baselines using an 11–parameter binary source fit. The fit parameters are still evolving and are not shown.

---

[2] http://darkstar.astro.washington.edu
[3] subscribe by mailing to macho@astro.washington.edu

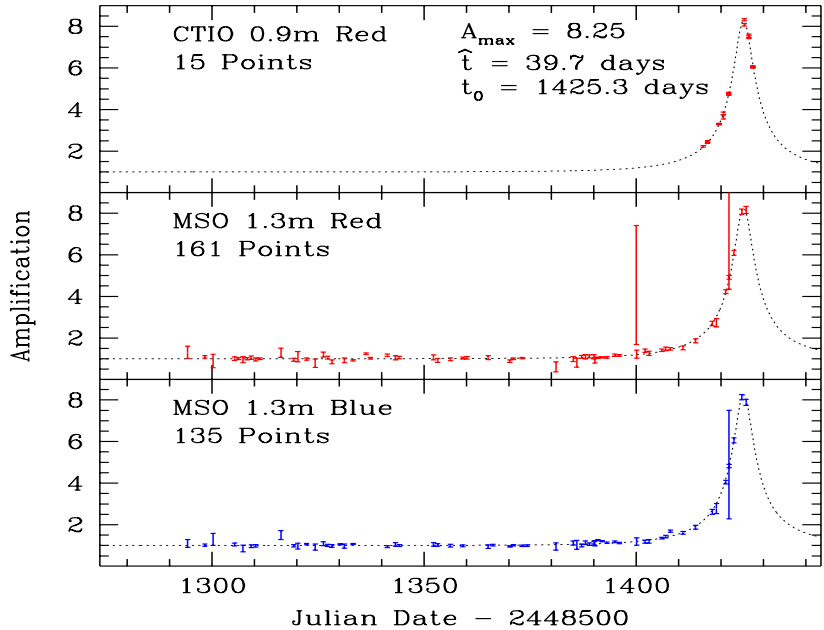

**Figure 2.** MACHO Object 21161_7671 (alert 95-26). The microlensing fit uses the three shape parameters described in Sect. 1 and an additional baseline parameter for each passband. An additional year of data from the '93 bulge season is used in the fit to better establish baselines but is not visible in this window. The CTIO data are preliminary and subject to change.

## 3. Conclusion

Both MACHO and OGLE collaborations have demonstrated the ability to distinguish with high reliability microlensing events in progress while discriminating against a very large background of variable stars. The spectroscopic followup of MACHO alert 94-1 [2, 4] has provided perhaps the most convincing case for detection of microlensing. More than 30 events have been detected in progress as of this writing. At least one of these events deviates significantly from the standard microlensing model. Development of more robust discrimination and more sensitive triggering should increase the detection rate. In closing, the ability to detect microlensing events in progress is crucial for such future microlensing endeavors as ground based planet searches [8] and microlensing parallax measurement from a satellite in solar orbit [6].

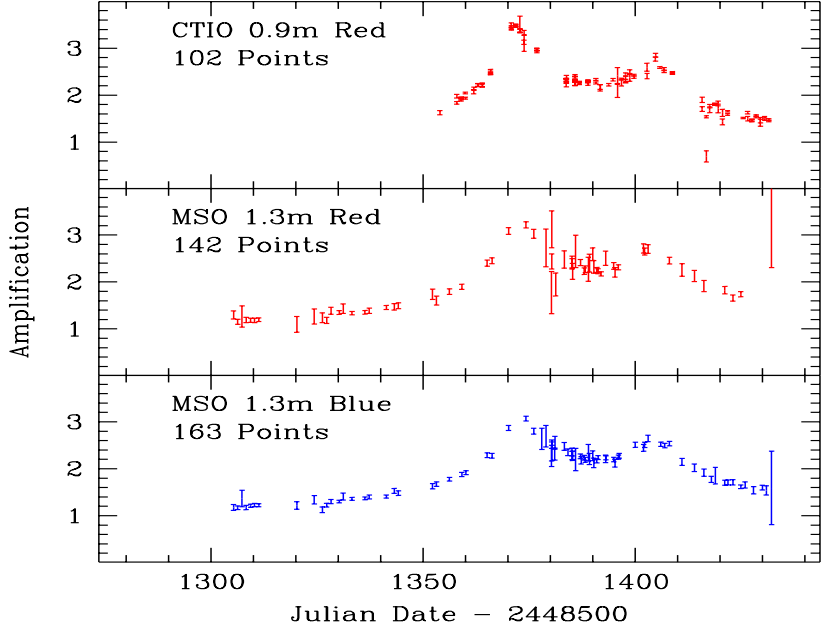

**Figure 3.** MACHO Object 21263_1213 (alert 95-12). There are a year of data from the '93 bulge season that have been used in the fit to provide accurate determinations of the quiescent luminosity but are not visible in this window. The CTIO data are preliminary and subject change.

## 4. Acknowledgments

We are very grateful for the skilled support given our project by S. Chan and the technical staff at the Mt. Stromlo Observatory. We especially thank J.D. Reynolds for the network software that has made this effort successful.